\documentclass[prl,showpacs,superscriptaddress,twocolumn]{revtex4}

\usepackage{amsmath,bm}
\usepackage{graphicx}
\usepackage{epstopdf}
\usepackage{subfigure}
\usepackage{epsfig}
\usepackage{amsmath,amssymb,amsfonts}
\usepackage{color}
\usepackage[utf8]{inputenc}
\usepackage{verbatim}
\usepackage{color}
\graphicspath{{figures/}} 
\begin{document}

\title{Jet charge in high energy nuclear collisions}

\date{\today  \hspace{1ex}}

\author{Shi-Yong Chen}
\affiliation{Key Laboratory of Quark \& Lepton Physics (MOE) and Institute of Particle Physics,
 Central China Normal University, Wuhan 430079, China}

\author{Ben-Wei Zhang}
\email{bwzhang@mail.ccnu.edu.cn}
\affiliation{Key Laboratory of Quark \& Lepton Physics (MOE) and Institute of Particle Physics,
 Central China Normal University, Wuhan 430079, China}

\author{Enke Wang}
\affiliation{ School of Physics and Telecommunication Engineering,
South China Normal University, Guangzhou 510006, China}
\affiliation{Key Laboratory of Quark \& Lepton Physics (MOE) and Institute of Particle Physics,
 Central China Normal University, Wuhan 430079, China}

\begin{abstract}
Averaged jet charge characterizes the electric charge for the initiating parton,
and provides a powerfull tool to distinguish quark jets from gluon jets. In this paper, we give the first prediction for the medium modification of averaged jet charge in heavy-ion collision at the LHC, where jet productions in p+p collisions are simulated by PYTHIA6, and parton energy loss in QGP are calculated with two Monte Carlo models of jet quenching: PYQUEN and JEWEL. We found that the distribution of averaged jet charge is significantly suppressed by initial state isospin effects due to the participants of neutrons with zero electric charge during nuclear collisions. Considerable enhancement of averaged jet charge in central Pb+Pb collisions relative to peripheral collisions is observed, since jet quenching effect is more pronounced in central collisions. Distinct feature of averaged jet charge between quark and gluon jets, together with the sensitivity of medium modifications on jet charge to flavor dependence of parton energy loss, could be very useful to discriminate the energy loss pattern between quark and gluon jets in heavy-ion collisions.
\end{abstract}

\maketitle
\section{Introduction}
\label{introduction}
One major purpose of relativistic heavy-ion collisions (HIC) performed at the RHIC~\cite{Adler:2003qi,Adams:2003kv,Adare:2008qa,Agakishiev:2011dc}
and the LHC~\cite{Aad:2010bu,Aamodt:2010jd,Chatrchyan:2011sx} is studying the formation and mapping the properties of the Quark-Gluon Plasma (QGP), a new kind of matter with de-confined quarks and gluons.  One optimal probe of the creation and properties of the QGP is the jet quenching, a phenomena that an energetic parton propagating in the dense QCD medium may suffer multiple scattering with other partons in medium and then loses a significant amount of its energy~\cite{Wang:1991xy,Gyulassy:2003mc,Qin:2015srf}. By indirectly quantifying how much the parton loses its energy in medium with theoretical estimation and experimental data of final-state observables at large transverse momentum such as leading hadron production we can infer the transport properties of the QCD medium created in HIC~\cite{Burke:2013yra,Chen:2010te,Chen:2011vt,Abelev:2014laa,Liu:2015vna,Dai:2017piq,Ma:2018swx,Xie:2019oxg}.

An essential part of jet quenching theory is the flavor dependence of parton energy loss, i.e. the difference energy loss patterns of quarks and gluons~\cite{Wang:1998bha,Gyulassy:2003mc}.  In most perturbative QCD models of jet quenching it is shown that an energetic gluon may lose more energy than quark, which gives $\Delta E_g/ \Delta E_q = (C_A/C_F)$ for asymptotic high-energy partons. However, in other models, the flavor pattern of energy loss may be different. For example, a hybrid strong/weak coupling model~\cite{Casalderrey-Solana:2014bpa} utilizes parton energy loss results in a strongly coupled plasma from the string theory with $\Delta E_g/ \Delta E_q = (C_A/C_F)^{1/3}$. The  difference of energy loss of quarks and gluons may change the parton fraction in HIC relative to that in elementary proton-proton collisions, and thus modify the hadron chemistry~\cite{Wang:1998bha,Liu:2006sf,Chen:2008vha,Dai:2015dxa,Dai:2017tuy}.  A deep understanding of flavor dependence of parton energy loss is indispensable to solve the baryon anomaly, which is still an open question
so far~\cite{Liu:2006sf,Chen:2008vha,Brodsky:2008qp}. We note another topic of favor dependence of parton energy loss is the mass dependence of heavy quark energy loss, and please see Refs.~\cite{Dokshitzer:2001zm, Zhang:2003wk, Djordjevic:2003qk,Sharma:2009hn,Cao:2013ita,Dong:2019unq} for more discussions on the mass hierarchy of jet quenching.

Since the start of heavy-ion programs with the unprecedented colliding energies available at the LHC, full jet observables in HIC have
attracted intense investigations both in theory and experiment~\cite{Vitev:2008rz, Vitev:2009rd, CasalderreySolana:2010eh,Young:2011qx,He:2011pd,ColemanSmith:2012vr,Neufeld:2010fj,Zapp:2012ak,Dai:2012am,Ma:2013pha, Senzel:2013dta, Casalderrey-Solana:2014bpa,Milhano:2015mng,Chang:2016gjp,Majumder:2014gda, Chen:2016cof, Chien:2016led, Apolinario:2017qay,Connors:2017ptx,Zhang:2018urd}. Jet observables could provide complementary information besides leading hadron productions and have been widely accepted as another excellent probe to investigate the
properties of QGP. It is of great interest to study jet observable which is sensitive to the pattern of gluon versus quark energy loss, which can help make further constrains of jet quenching mechanism by utilizing the large amount data on jet measurements at the LHC and also at the RHIC. One of such kind of jet observables is the averaged jet charge~\cite{Field:1977fa,Krohn:2012fg,Waalewijn:2012sv}, which gives the electric charge distribution in a reconstructed jet.
It is expected that the measurement of jet charge in HIC may shed an insight of flavor dependence of jet quenching~\cite{Chen:2017uqx}.

In this paper, we present the first numerical calculations of medium modification on averaged jet charge in Pb+Pb collisions at the LHC.
PYHHIA6~\cite{Sjostrand:2006za} is used to simulate particle productions in p+p collisions. PYQUEN and JEWEL are used to simulate parton energy loss in Pb+Pb collisions.
It is found that as compared to p+p collisions, averaged jet charge is significantly suppressed due to the participating of neutron in heavy-ion collisions. However, with the input that gluon loses more energy than quark, the fraction of quark at large transverse $p_T$ will be increased to large extent,  it is shown that $R_{CP}$, the ratio of averaged jet charge in central collisions to that in peripheral collisions, should be larger than unit over the whole range of jet transverse momentum. We demonstrate that the behavior of central-to-peripheral ratio is especially sensitive to
the difference of the gluon and quark jet-medium interaction strength.

\section{Theoretical framework}
In this section, we discuss our analysis framework of averaged jet charge from p+p to Pb+Pb collisions.
The momentum-weighted jet charge is defined as~\cite{Krohn:2012fg,Waalewijn:2012sv}:

\begin{eqnarray}
Q^{\kappa}=\sum_{i\in jet}z_h^{\kappa}Q_{i}
\label{TAB}
\end{eqnarray}
with $z_h={p^{i}_{T}}/{p^{\text{jet}}_{T}}$. Here $p^{i}_{T}$ indicates the transverse momentum of the hadron $i$ inside jet, $p^{\text{jet}}_{T}$ the transverse momentum of the jet,
$Q_{i}$ is the electric charge of the hadron.
A power parameter $\kappa$  which satisfies $\kappa\in(0.1,1)$ is used to adjust
the contribution bias of jet constituents with different transverse momentum.

\begin{figure}[!htb]
\centerline{
\includegraphics[width = 95mm]{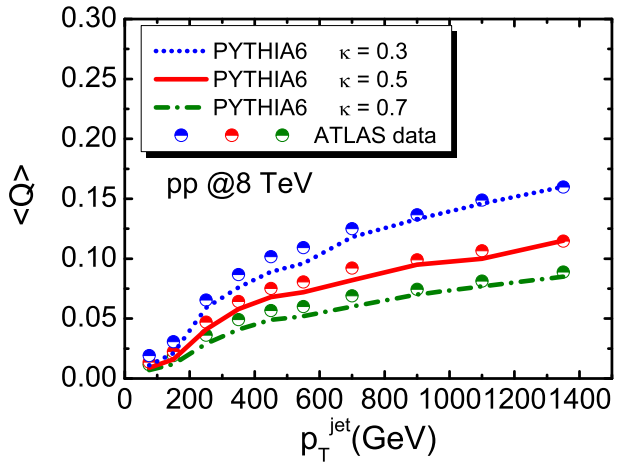}
}
\caption{The average value of the jet charge distribution for the leading jet
in dijet events, as a function of the jet transverse momentum at $\sqrt{s}=8.0$~TeV in p+p collisions, as compared with ATLAS data.}
\label{fig:ATLAS}
\end{figure}

It has been shown that
the energy and and jet-size dependence of moments of jet-charge distributions in p+p can be
calculated in perturbative QCD~\cite{Krohn:2012fg,Waalewijn:2012sv}.
In this paper, we employ Monte Carlo event generater PYTHIA
with Perugia 2012 tune~\cite{Skands:2010ak}
to simulate particle production in p+p collisions which gives us more leverage power to impose different kinematic cuts, and FastJet package~\cite{Cacciari:2008gp} is used to construct final state jets for jet charge calculation.
In ATLAS Collaboration measurement, jets are reconstructed by the anti-$k_{T}$
algorithm with radius $R=0.4$,
events are required to include at least
two jets with $p_{T}>50$~GeV in central rapidity regiom $\left|\eta_{1,2}\right|<2.1$. Only the leading two jets with $p^{\text{leading}}_{T}/p^{\text{subleading}}_{T}<1.5$ are used
in the jet charge calculations.
Shown in  Fig.~\ref{fig:ATLAS} are our results for jet charge at $\sqrt{s}=8$~TeV in p+p collisions from {PYTHIA+FastJet}
compared with ATLAS recent data~\cite{Aad:2015cua}.
One can find that our MC results give very nice descriptions for experimental data. We may calculate jet charge
at $\sqrt{s}=2.76$~TeV p+p collisions by using the same method, which provides a good baseline for the calculation in heavy-ion collisions.

\begin{figure}[!htb]
\centerline{
\includegraphics[width = 95mm]{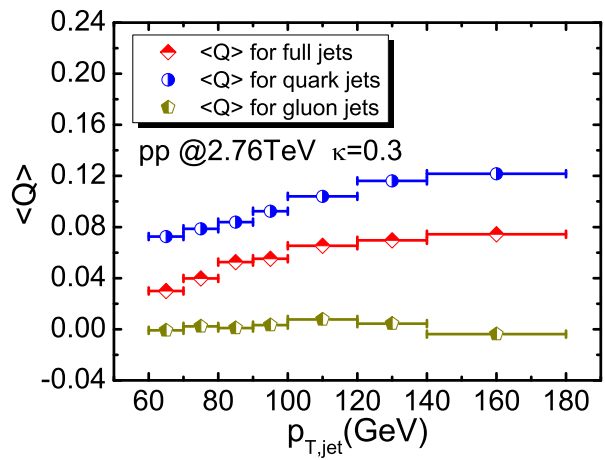}
}
\caption{The jet charge for quark and gluon jet at $\sqrt{s}=2.76$~TeV in p+p collisions.}
\label{fig:qg}
\end{figure}

We show the jet charge for quark and gluon jets respectively in Fig.~\ref{fig:qg}.
Distinct feature of jet charge between quark and gluon jets can be observed.  The value of averaged full jet charge is the combination of quark and gluon jets with their relative fractions. Because the fraction of quark jets goes up with increasing transverse momentum. The  averaged jet charge may increase gradually with jet transverse momentum $p_T^{\text{jet}}$, as shown in Fig.~\ref{fig:ATLAS} and Fig.~\ref{fig:qg}. Moreover, even for pure quark jets, the value of averaged jet charge is increased. We note that hadronization effect decreases with higher jet $p_T^{\text{jet}}$, thus the out-of-cone contributions of electric and energy for high $p_T^{\text{jet}}$ jet should be suppressed  as compared to that at the low $p_T^{\text{jet}}$.

Now, we move to the calculations of jet charge in heavy-ion collisions at the LHC,
where both initial-state cold nuclear matter (CNM) effects and final-state hot nuclear matter
effects should be taken into account. For CNM effects, EPPS16 parameterization set~\cite{Eskola:2016oht} has been utilized in our calculations. It is noticed that the total value of averaged jet charge
 comes originally from the electric charge of colliding participants. In Pb+Pb collisions at the LHC, besides proton-proton collisions, proton-neutron and neutron-neutron collisions
may also happen. With the approximation that the cross sections of these three types of collisions are almost the same, the probability of these collisions
is determined by the number fractions of colliding participants. Eventually, the averaged jet charge in Pb+Pb can be given as:

\begin{equation}
\begin{split}
\langle Q^{\kappa}_{PbPb} \rangle = {} &\left(\frac{N_{p}}{208}\right)^{2}\langle Q^{\kappa}_{pp}\rangle+\left(\frac{N_{p}}{208}\frac{N_{n}}{208}\right)\langle Q^{\kappa}_{pn}\rangle {} \\
&+\left(\frac{N_{n}}{208}\right)^{2}\langle Q^{\kappa}_{nn}\rangle
\end{split}
\label{ppnn}
\end{equation}
where $N_{p}$ and $N_{n}$ represent the number of proton and neutron in Lead nucleus respectively.  With Eq.~(\ref{ppnn}), one can calculate the averaged jet charge in
Pb+Pb collisions without the final-state jet quenching effect, as shown in Fig.~\ref{fig:ppnn}. The CNM modification ratio is expressed as:

\begin{figure}[!htb]
\centerline{
\includegraphics[width = 95mm]{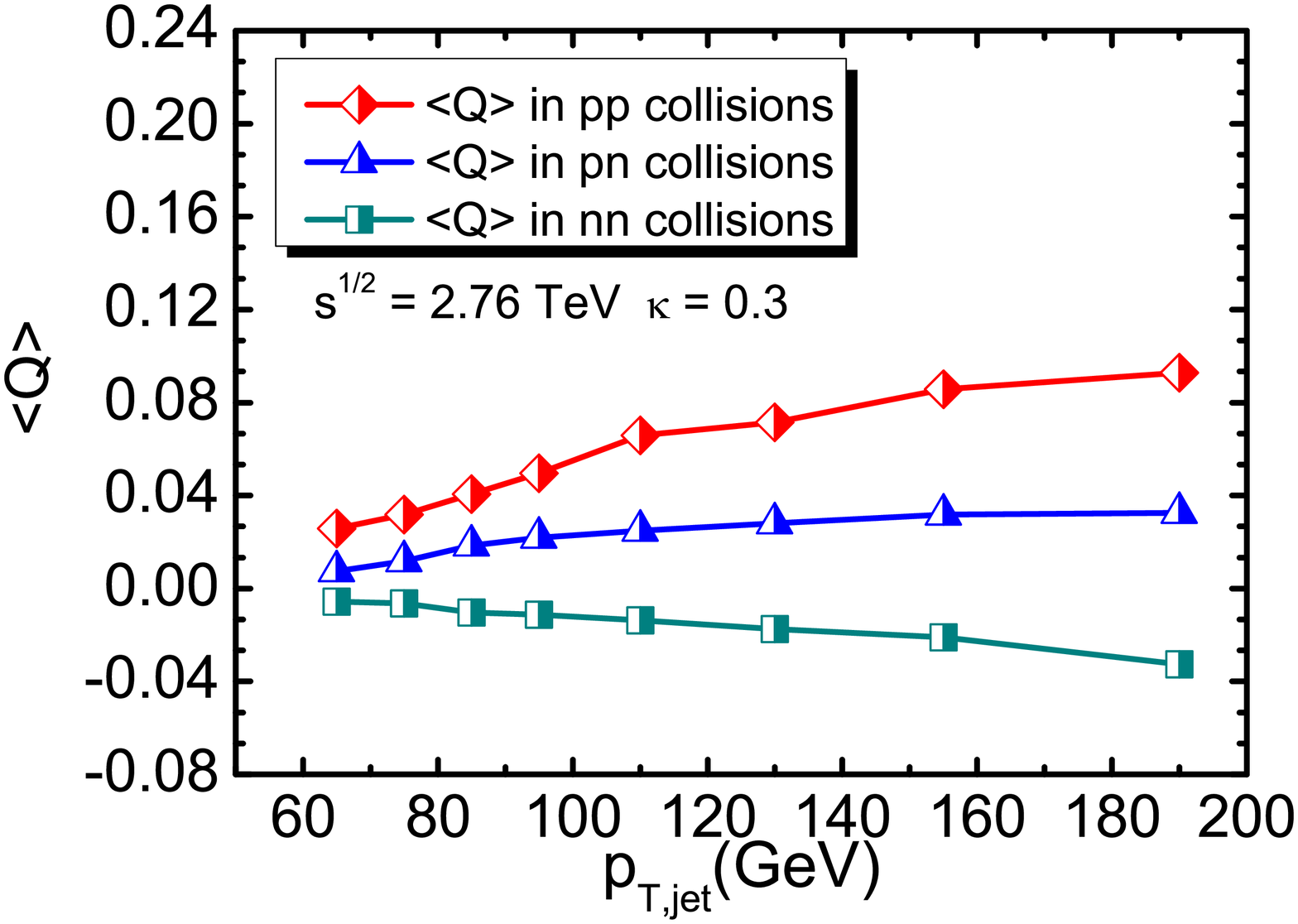}
}
\caption{The average value of the jet charge distribution for the leading jet
in dijet events as a function of the jet transverse momentum in pp, pn and nn collisions.}
\label{fig:ppnn}
\end{figure}

\begin{eqnarray}
R_{\text{CNM}}=\frac{\langle Q_{AA}^{\text{CNM}} \rangle}{\langle Q_{pp} \rangle}
\label{cnm}
\end{eqnarray}

\begin{figure}[!htb]
\centerline{
\includegraphics[width = 95mm]{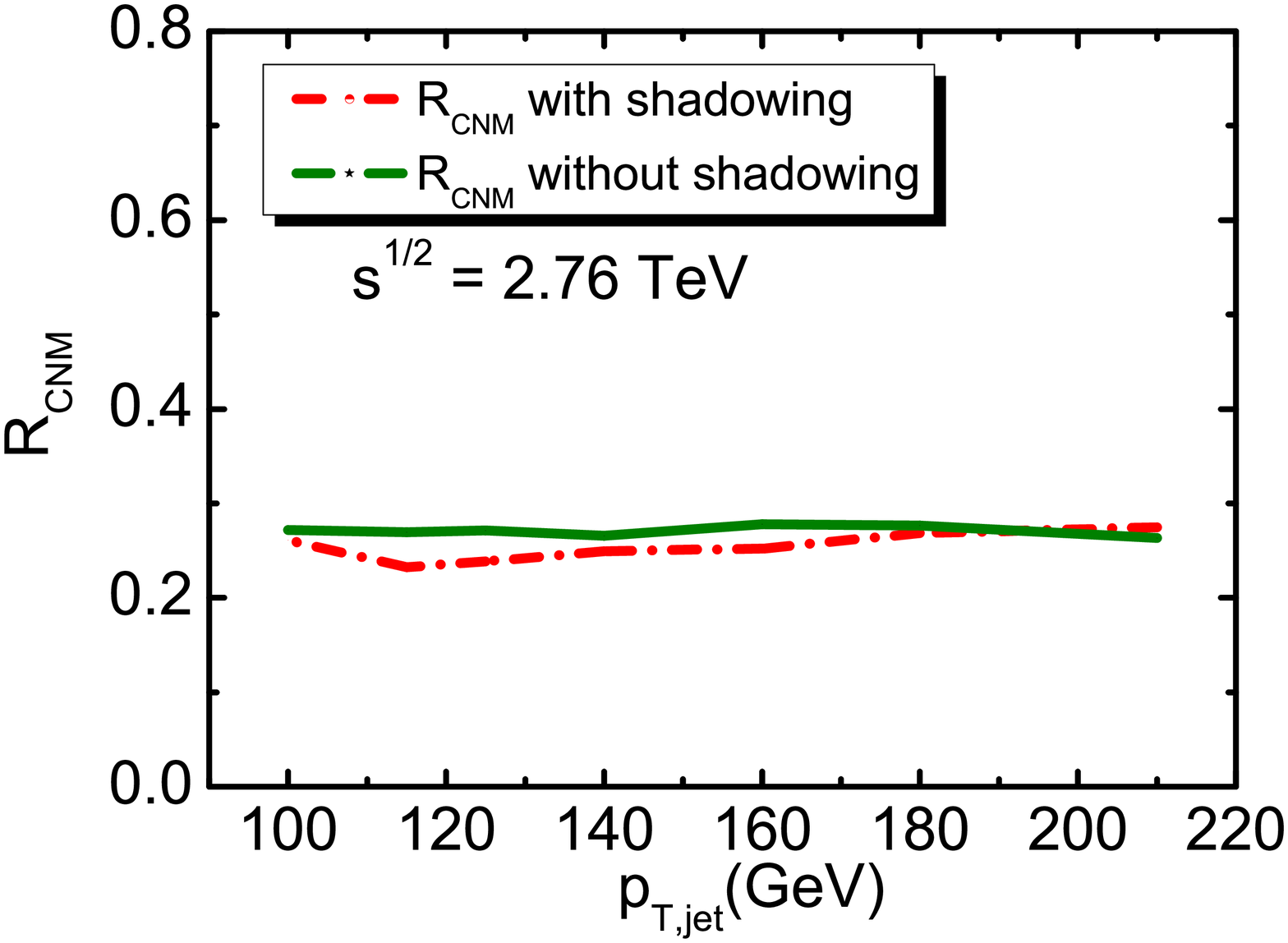}
}
\caption{Cold nuclear modification ratio for averaged jet charge.}
\label{fig:CNM}
\end{figure}

The CNM modification ratio as a function
of $p^{\text jet}_{T}$ is illustrated in Fig.~\ref{fig:CNM}. As we can see,  averaged jet charge with CNM effects is significant suppressed relative to that in p+p due to the participants of neutrons
with zero electric charge during nuclear collisions.

Previous studies have shown that energetic partons should lose energy when traversing QCD medium due
to elastic and inelastic processes.
In this paper we employ PYQUEN and JEWEL models to simulate parton energy loss in the QGP medium.
PYQUEN~\cite{Lokhtin:2005px,Lokhtin:2006qm,Lokhtin:2011qq} is one of the Monte Carlo event generator
of jet quenching and built as a modification
of jet events obtained for hadronic collisions with
PYTHIA 6.4. The details of the used physics model
and simulation procedure can be found in~\cite{Lokhtin:2005px}.
The model has include both
radiative and collisional energy loss of hard partons, as well as the
realistic nuclear geometry. To investigate jet observables, another key element of PYQUEN is the angular spectrum of medium-induced gluon
radiation. It has been found that the "wide-angle" radiation scenario could provide better description for experimental measurements on full jets observables in PYQUEN model~\cite{Lokhtin:2014vda}, which is also adopted in the calculations in this paper.

JEWEL event generator provided good description of jet evolution in the QCD medium created in ultra-relativistic heavy-ion collisions, which has been well tested for a large set of jet
quenching measurements, including jet $R_{AA}$,
dijet asymmetry $A_{J}$, bosons tagged jet as well as jet substructure observables
which are more sensitive to jet-medium interaction~\cite{Zapp:2008gi,Zapp:2011ya,Milhano:2015mng,Zapp:2012ak,KunnawalkamElayavalli:2016ttl,KunnawalkamElayavalli:2017hxo}.
In JEWEL the recoiling scattering centers can be traced and in
principle they could be subjected to further interactions. To deal with medium response, JEWEL provide one option
that thermal partons are kept recoiling against interactions with the jet partons
and then fragment into hadrons together with the jet partons. This option is used in the following simulations.

\section{Results and Discussion}

Based the framework established in the last section, we calculate the average value of the jet charge distribution as function of the jet transverse momentum in HIC at the LHC. The CNM effects modified
partonic jets are generated by PYTHIA+EPPS16 with Perugia 2012 tune~\cite{Skands:2010ak}, and suffered jet quenching with
simulations of PYQUEN and JEWEL models, and then fragmented into final-state hadrons.
Jets are reconstructed by using the anti-$k_{t}$ jet finding algorithm~\cite{Cacciari:2008gp} with cone size $R=0.4$ implemented in
FastJet. Events are required to have two leading jets with $p^{\text{leading}}_{T}>100$~GeV, $p^{\text{subleading}}_{T}>50$~GeV, $|\eta|<2.1$.
Where $p^{\text{leading}}_{T}$ and $p^{\text{subleading}}_{T}$ are the transverse momentum of the two leading jets.
To demonstrate the nuclear modification
of averaged jet charge in Pb+Pb collisions,
the nuclear modification factor is given by:

\begin{eqnarray}
R_{AA}=\frac{\langle Q_{AA} \rangle}{\langle Q_{pp} \rangle}
\label{raa}
\end{eqnarray}

\begin{figure}[!htb]
\centerline{
\includegraphics[width = 95mm]{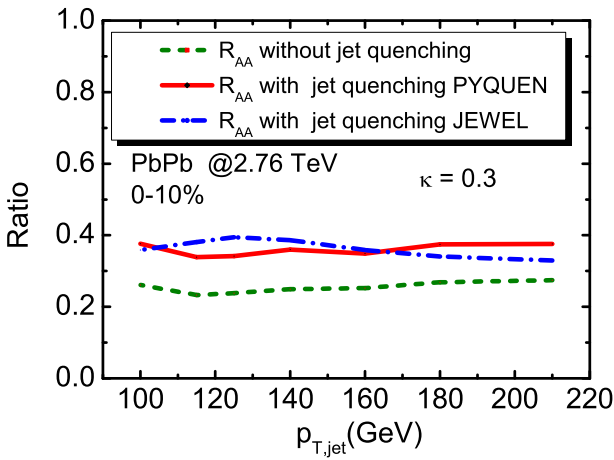}
}
\caption{The nuclear modification factor for averaged jet charge and their comparison with cold nuclear modification ratio in Pb+Pb collisions with $\sqrt{s_{NN}}=2.76$~TeV.}
\label{fig:2.76}
\end{figure}

Shown in Fig.~\ref{fig:2.76} are our predictions for jet charge $R_{AA}$ and their comparison with cold nuclear modification ratio in central 0-10~\%) Pb+Pb collisions.
One can observe the nuclear modification factor for jet charge may increase considerably by jet quenching effect for both PYQUEN and JEWEL. To give clearer description of this phenomenon and suppress the initial-state isospin effect (the electronic charge difference between protons and neutrons), we study the central-to-peripheral ratio
of averaged jet charge in Pb+Pb collisions $R_{CP}$, while the jet quenching effect is much stronger in central collisions than that in peripheral collisions. The central-to-peripheral ratio $R_{CP}$ is defined as:

\begin{eqnarray}
R_{CP}=\frac{\langle Q_{AA}^{\text{central}} \rangle}{\langle Q_{AA}^{\text{peripheral}} \rangle}
\label{raa}
\end{eqnarray}

\begin{figure}[!htb]
\centerline{
\includegraphics[width = 95mm]{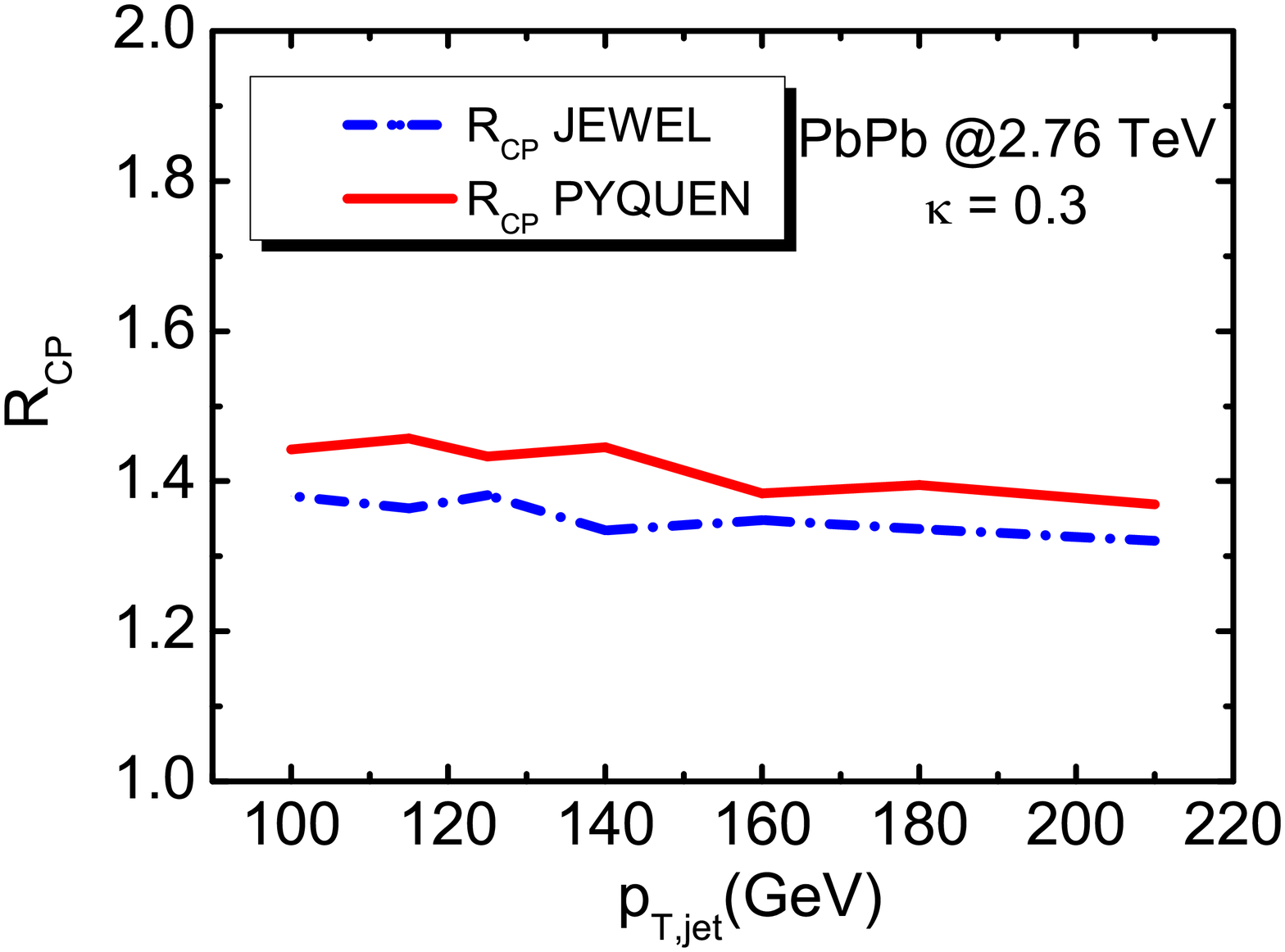}
}
\caption{The central-to-peripheral ratio $R_{CP}$
of averaged jet charge in Pb+Pb collisions with $\sqrt{s_{NN}}=2.76$~TeV.}
\label{fig:rcp}
\end{figure}

In Fig.~\ref{fig:rcp}, we plot the $R_{CP}$ as function of jet transverse momentum for averaged jet charge from $0-10$~\% centrality ($b\in(0,3.478)$) and $60-80$~\% centrality ($b\in(12.05,13.91)$) Pb+Pb collisions.
It is shown that  $R_{CP}$ is significantly larger than  in the whole range of jet transverse momentum due to jet quenching effect both from PYQUEN and JEWEL. We emphasize the behavior that $R_{CP}>1 $ for jet charge comes mainly from the increasing fraction of quark jet with $p_T^{\text{jet}}$ due to jet quenching effect in both PYQUEN and JEWEL models, where $\Delta E_{g}/\Delta E_{q}=C_{A}/C_{F}=9/4$ has been implemented. In these models because gluon loses more energy than quarks, the quark fraction will increase and gluon fraction will decrease. However as we discussed before, the averaged jet charge for gluon jet approaches nearly zero. Therefore jet quenching model with gluon losing much more energy than quark naturally result in a larger than unit $R_{CP}$.

\begin{figure}[!htb]
\includegraphics[width = 95mm]{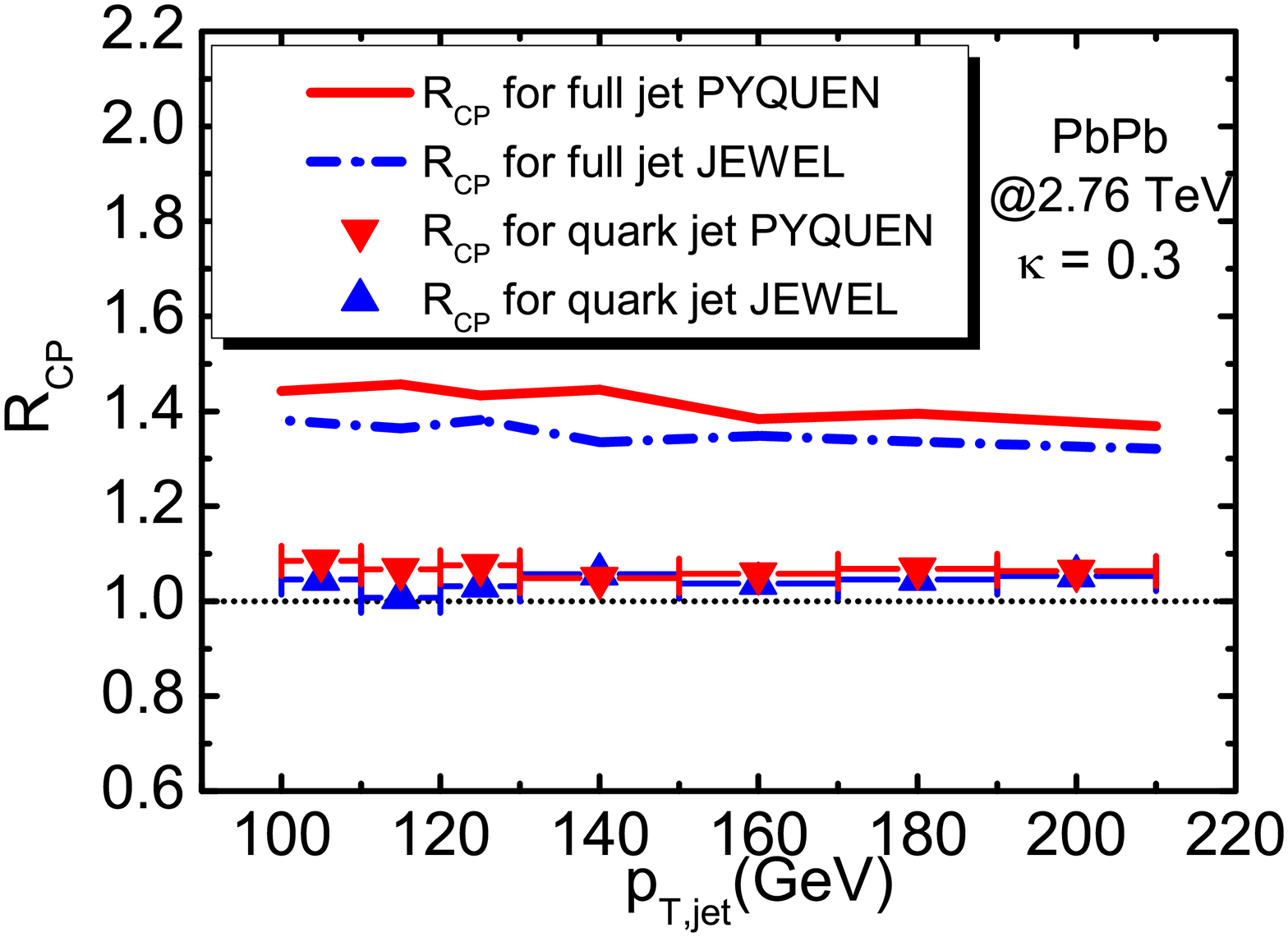}
\caption{ $R_{CP}$ for quark jet charge and and full jet charge in Pb+Pb at the LHC.}
\label{fig:R-q-jet}
\end{figure}

To clarify this, we study numerically the central-to-peripheral ratio ($R^{q,\text{jet}}_{CP}$) for averaged quark jet charge. The value of $R^{q,\text jet}_{CP}$ and their comparison with full jets are shown in Fig.~\ref{fig:R-q-jet}.  We observe that
$R^{q,\text jet}_{CP}\sim 1$, which implies that jet quenching effect has rather modest impact of pure quark jet charge; the difference between full charge in central and peripheral collisions mainly results from the decreasing fraction of gluon jet with energy loss effect in the hot QCD medium.

\begin{figure}[!htb]
\includegraphics[width = 95mm]{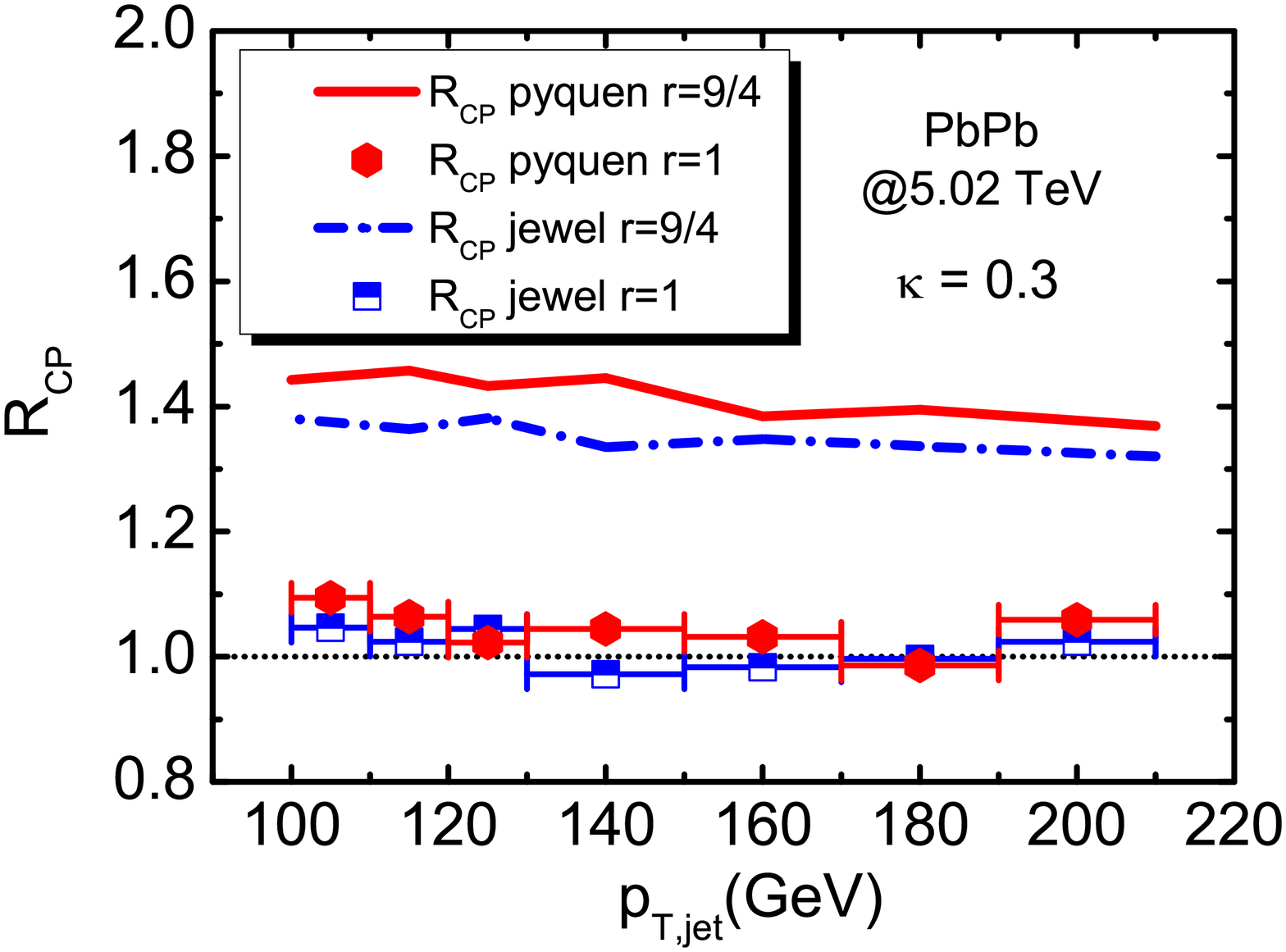}
\caption{$R_{CP}$ for the averaged jet charge in jet quenching models with different flavor patterns.}
\label{fig:f1}
\end{figure}

\begin{figure}[!htb]
\includegraphics[width = 95mm]{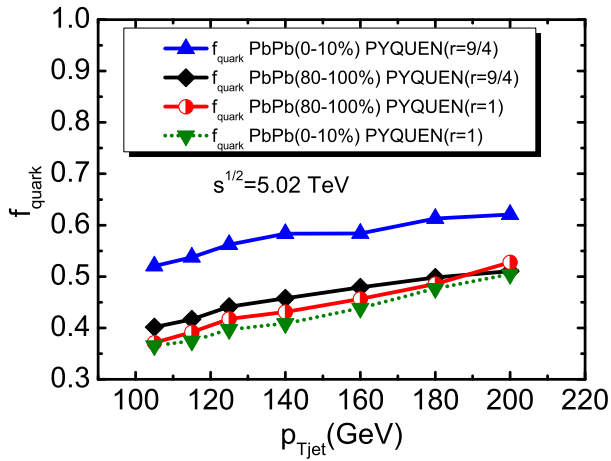}
\caption{The fraction of quark jets $f_{\text{quark}}$ in jet quenching models with different flavor patterns.}
\label{fig:quark-ratio}
\end{figure}

To further investigate the sensitivity of medium modifications of jet charge in HIC to the flavor dependence of parton energy loss,
we calculate $R_{CP}$  in jet quenching models with different values of:

\begin{eqnarray}
r=\Delta E_{g}/\Delta E_{q}
\label{r}
\end{eqnarray}

We consider two scenarios: Scenario (I) where we have $r=9/4$ for conventional jet quenching model; Scenario (II) with  $r=1$. 
We note that the different choices of $r$ may alter hadron chemistry such as
$p^{+}/\pi^{+}$ ratio
in A+A collisions~\cite{Chen:2008vha}.

In Fig.~\ref{fig:f1} and Fig.~\ref{fig:quark-ratio} we illustrate $R_{CP}$ for the average jet charge and quark jet fraction $f_{\text{quark}}$ in these two scenarios of flavor dependence of parton energy loss model $r=9/4$
and $r=1$ respectively. It is seen that in Scenario (II) with $r=1$ the central-to-peripheral ratio $R_{CP}$ is about $1$, and a very small difference of quark jet fraction between central and peripheral Pb+Pb collisions is observed. This striking distinction between results in Scenario (I) and Scenario (II) demonstrate convincingly the sensitivity of jet charge to the energy loss pattern between quark and gluon jets in heavy-ion collisions. By measuring jet charge in HIC and comparing the data with theoretical results from different jet quenching model we can make stringent constraints on jet quenching mechanism.

\section{Summary and Conclusions}

We present the first numerical results for the nuclear modification of averaged jet charge in HIC at the LHC. In p+p collisions, PYTHIA6 is used to simulate final state hadron productions.
It is shown that jet charge for quark goes up with increasing jet transverse momentum, while  gluon jet charge approximately is zero in the whole range of $p_{T}$ because gluon carries no electric charge. In Pb+Pb collisions, EPPS16 parameter set is used to investigate cold nuclear effects. We employ MC event generator PYQUEN and JEWEL to simulated parton energy loss in QCD medium.
We found that during nuclear collisions, averaged jet charge with cold nuclear matter effects is significantly suppressed by isospin effects. That is because neutrons with zero electric
charge accounting for a large proportion in the collisions. To study medium modification for jet charge by jet quenching effects and suppress isospin effect, we calculate central-to-peripheral ratio
for jet charge in Pb+Pb collisions.
The value of averaged jet charge in central Pb+Pb collisions is enhanced relative to peripheral collisions, since jet quenching effect is much stronger in central collisions.
In conventional jet quenching calculations, a fast gluon will lose more energy in the QGP than a fast quark due to its large color-charge($\Delta E_{g}/\Delta E_{q}=C_{A}/C_{F}=9/4$), more quarks with electric charge may survive in central collisions as compared with that in peripheral Pb+Pb collisions. The fraction of quark jet should be increased, which results in the larger jet charge in central collisions than that in peripheral reactions. We found the central-to-peripheral ratio of averaged jet charge is especially sensitive to flavor dependence of parton energy loss, which may provide a very powerful tool to constrain the energy loss pattern between quark and gluon jets in heavy-ion collisions.

\vspace*{.3cm}

{\it Acknowledgments:} This research is supported in part by the NSFC of China with Project Nos. 11435004.


\vspace*{-.6cm}


\begin{thebibliography}{99}





\bibitem{Adler:2003qi}
  S.~S.~Adler {\it et al.}  [PHENIX Collaboration],
  Phys.\ Rev.\ Lett.\  {\bf 91}, 072301 (2003).

\bibitem{Adams:2003kv}
  J.~Adams {\it et al.}  [STAR Collaboration],
  Phys.\ Rev.\ Lett.\  {\bf 91}, 172302 (2003).

\bibitem{Adare:2008qa}
  A.~Adare {\it et al.} [PHENIX Collaboration],
  Phys.\ Rev.\ Lett.\  {\bf 101}, 232301 (2008).

\bibitem{Agakishiev:2011dc}
  G.~Agakishiev {\it et al.} [STAR Collaboration],
  Phys.\ Rev.\ Lett.\  {\bf 108}, 072302 (2012).



\bibitem{Aad:2010bu}
  G.~Aad {\it et al.}  [ATLAS Collaboration],
  Phys.\ Rev.\ Lett.\  {\bf 105}, 252303 (2010).

\bibitem{Aamodt:2010jd}
  K.~Aamodt {\it et al.}  [ALICE Collaboration],
  Phys.\ Lett.\ B {\bf 696}, 30 (2011).

\bibitem{Chatrchyan:2011sx}
  S.~Chatrchyan {\it et al.}  [CMS Collaboration],
  Phys.\ Rev.\ C {\bf 84}, 024906 (2011).

\bibitem{Wang:1991xy}
  X.~N.~Wang and M.~Gyulassy,
  Phys.\ Rev.\ Lett.\  {\bf 68}, 1480 (1992).

\bibitem{Gyulassy:2003mc}
  M.~Gyulassy, I.~Vitev, X.~N.~Wang and B.~W.~Zhang,
  In *Hwa, R.C. (ed.) et al.: Quark gluon plasma* 123-191
  [nucl-th/0302077].

\bibitem{Qin:2015srf}
  G.~Y.~Qin and X.~N.~Wang,
  Int.\ J.\ Mod.\ Phys.\ E {\bf 24}, no. 11, 1530014 (2015).

\bibitem{Burke:2013yra}
  K.~M.~Burke {\it et al.} [JET Collaboration],
  Phys.\ Rev.\ C {\bf 90}, no. 1, 014909 (2014).

\bibitem{Chen:2010te}
  X.~Chen, C.~Greiner, E.~Wang, X.~N.~Wang and Z.~Xu,
  Phys.\ Rev.\ C {\bf 81}, 064908 (2010).

\bibitem{Chen:2011vt}
  X.~Chen, T.~Hirano, E.~Wang, X.~N.~Wang and H.~Zhang,
  Phys.\ Rev.\ C {\bf 84}, 034902 (2011).

\bibitem{Abelev:2014laa}
  B.~B.~Abelev {\it et al.} [ALICE Collaboration],
  Phys.\ Lett.\ B {\bf 736}, 196 (2014).

\bibitem{Liu:2015vna}
  Z.~Q.~Liu, H.~Zhang, B.~W.~Zhang and E.~Wang,
  Eur.\ Phys.\ J.\ C {\bf 76}, no. 1, 20 (2016).

\bibitem{Dai:2017piq}
  W.~Dai, X.~F.~Chen, B.~W.~Zhang, H.~Z.~Zhang, E.~Wang,
  Eur.\ Phys.\ J.\ C {\bf 77}, no. 8, 571 (2017).

\bibitem{Ma:2018swx}
  G.~Y.~Ma, W.~Dai, B.~W.~Zhang and E.~K.~Wang,
  Eur.\ Phys.\ J.\ C {\bf 79}, no. 6, 518 (2019).


\bibitem{Xie:2019oxg}
  M.~Xie, S.~Y.~Wei, G.~Y.~Qin and H.~Z.~Zhang,
  Eur.\ Phys.\ J.\ C {\bf 79}, no. 7, 589 (2019).


\bibitem{Wang:1998bha}
  X.~N.~Wang,
  Phys.\ Rev.\ C {\bf 58}, 2321 (1998).

\bibitem{Casalderrey-Solana:2014bpa}
  J.~Casalderrey-Solana, D.~C.~Gulhan, J.~G.~Milhano, D.~Pablos and K.~Rajagopal,
  JHEP {\bf 1410}, 019 (2014);
  Erratum: [JHEP {\bf 1509}, 175 (2015)].

\bibitem{Liu:2006sf}
  W.~Liu, C.~M.~Ko and B.~W.~Zhang,
  Phys.\ Rev.\ C {\bf 75}, 051901 (2007).



\bibitem{Chen:2008vha}
  X.~Chen, H.~Zhang, B.~W.~Zhang and E.~Wang,
  J.\ Phys.\  {\bf 37}, 015004 (2010).

\bibitem{Dai:2015dxa}
  W.~Dai, X.~F.~Chen, B.~W.~Zhang and E.~Wang,
  Phys.\ Lett.\ B {\bf 750}, 390 (2015).

\bibitem{Dai:2017tuy}
  W.~Dai, B.~W.~Zhang and E.~Wang,
   Phys.\ Rev.\ C {\bf 98}, 024901 (2018).


\bibitem{Brodsky:2008qp}
  S.~J.~Brodsky and A.~Sickles,
  Phys.\ Lett.\ B {\bf 668}, 111 (2008).

\bibitem{Dokshitzer:2001zm}
  Y.~L.~Dokshitzer and D.~E.~Kharzeev,
  Phys.\ Lett.\ B {\bf 519}, 199 (2001).

\bibitem{Zhang:2003wk}
  B.~W.~Zhang, E.~Wang and X.~N.~Wang,
  Phys.\ Rev.\ Lett.\  {\bf 93}, 072301 (2004).

\bibitem{Djordjevic:2003qk}
  M.~Djordjevic and M.~Gyulassy,
  Phys.\ Lett.\ B {\bf 560}, 37 (2003).

\bibitem{Sharma:2009hn}
  R.~Sharma, I.~Vitev and B.~W.~Zhang,
  Phys.\ Rev.\ C {\bf 80}, 054902 (2009).

\bibitem{Cao:2013ita}
  S.~Cao, G.~Y.~Qin and S.~A.~Bass,
  Phys.\ Rev.\ C {\bf 88}, 044907 (2013).


\bibitem{Dong:2019unq}
  X.~Dong and V.~Greco,
  Prog.\ Part.\ Nucl.\ Phys.\  {\bf 104}, 97 (2019).


\bibitem{Vitev:2008rz}
  I.~Vitev, S.~Wicks and B.~W.~Zhang,
  JHEP {\bf 0811}, 093 (2008).

\bibitem{Vitev:2009rd}
  I.~Vitev and B.~W.~Zhang,
  Phys.\ Rev.\ Lett.\  {\bf 104}, 132001 (2010).

\bibitem{CasalderreySolana:2010eh}
  J.~Casalderrey-Solana, J.~G.~Milhano and U.~A.~Wiedemann,
  J.\ Phys.\ G {\bf 38}, 035006 (2011).



\bibitem{Young:2011qx}
  C.~Young, B.~Schenke, S.~Jeon and C.~Gale,
  Phys.\ Rev.\ C {\bf 84}, 024907 (2011).

\bibitem{He:2011pd}
  Y.~He, I.~Vitev and B.~W.~Zhang,
  Phys.\ Lett.\ B {\bf 713}, 224 (2012).


 \bibitem{ColemanSmith:2012vr}
  C.~E.~Coleman-Smith and B.~Muller,
  Phys.\ Rev.\ C {\bf 86}, 054901 (2012).

\bibitem{Neufeld:2010fj}
  R.~B.~Neufeld, I.~Vitev and B.-W.~Zhang,
  Phys.\ Rev.\ C {\bf 83}, 034902 (2011).

\bibitem{Zapp:2012ak}
  K.~C.~Zapp, F.~Krauss and U.~A.~Wiedemann,
  JHEP {\bf 1303}, 080 (2013).

\bibitem{Dai:2012am}
  W.~Dai, I.~Vitev and B.~W.~Zhang,
  Phys.\ Rev.\ Lett.\  {\bf 110}, 142001 (2013).


\bibitem{Ma:2013pha}
  G.~L.~Ma,
  Phys.\ Rev.\ C {\bf 87}, no. 6, 064901 (2013).


\bibitem{Senzel:2013dta}
  F.~Senzel, O.~Fochler, J.~Uphoff, Z.~Xu and C.~Greiner,
  J.\ Phys.\ G {\bf 42}, no. 11, 115104 (2015).




\bibitem{Majumder:2014gda}
  A.~Majumder and J.~Putschke,
  Phys.\ Rev.\ C {\bf 93}, no. 5, 054909 (2016).


\bibitem{Milhano:2015mng}
  J.~G.~Milhano and K.~C.~Zapp,
  Eur.\ Phys.\ J.\ C {\bf 76}, no. 5, 288 (2016).

\bibitem{Chang:2016gjp}
  N.~B.~Chang and G.~Y.~Qin,
  Phys.\ Rev.\ C {\bf 94}, no. 2, 024902 (2016).

\bibitem{Chen:2016cof}
  L.~Chen, G.~Y.~Qin, S.~Y.~Wei, B.~W.~Xiao and H.~Z.~Zhang,
  Phys.\ Lett.\ B {\bf 782}, 773 (2018).


\bibitem{Chien:2016led}
  Y.~T.~Chien and I.~Vitev,
  Phys.\ Rev.\ Lett.\  {\bf 119}, no. 11, 112301 (2017).


\bibitem{Apolinario:2017qay}
  L.~Apolinario, J.~G.~Milhano, M.~Ploskon and X.~Zhang,
  Eur.\ Phys.\ J.\ C {\bf 78}, no. 6, 529 (2018)

\bibitem{Connors:2017ptx}
  M.~Connors, C.~Nattrass, R.~Reed and S.~Salur,
  Rev.\ Mod.\ Phys.\  {\bf 90}, 025005 (2018)

\bibitem{Zhang:2018urd}
  S.~L.~Zhang, T.~Luo, X.~N.~Wang and B.~W.~Zhang,
  Phys.\ Rev.\ C {\bf 98}, 021901 (2018).

\bibitem{Field:1977fa}
  R.~D.~Field and R.~P.~Feynman,
  Nucl.\ Phys.\ B {\bf 136}, 1 (1978).



\bibitem{Krohn:2012fg}
  D.~Krohn, M.~D.~Schwartz, T.~Lin and W.~J.~Waalewijn,
  Phys.\ Rev.\ Lett.\  {\bf 110}, no. 21, 212001 (2013)


\bibitem{Waalewijn:2012sv}
  W.~J.~Waalewijn,
  Phys.\ Rev.\ D {\bf 86}, 094030 (2012)


 \bibitem{Chen:2017uqx}
  S.~Y.~Chen, B.~W.~Zhang and E.~K.~Wang,
  Nucl.\ Part.\ Phys.\ Proc.\  {\bf 289-290}, 448 (2017)
  [arXiv:1701.01308 [nucl-th]].


\bibitem{Sjostrand:2006za}
  T.~Sjostrand, S.~Mrenna and P.~Z.~Skands,
  JHEP {\bf 0605}, 026 (2006)

\bibitem{Skands:2010ak}
  P.~Z.~Skands,
  Phys.\ Rev.\ D {\bf 82}, 074018 (2010)

\bibitem{Cacciari:2008gp}
  M.~Cacciari, G.~P.~Salam and G.~Soyez,
  JHEP {\bf 0804}, 063 (2008).

\bibitem{Aad:2015cua}
  G.~Aad {\it et al.} [ATLAS Collaboration],
  arXiv:1509.05190 [hep-ex].

\bibitem{Eskola:2016oht}
  K.~J.~Eskola, P.~Paakkinen, H.~Paukkunen and C.~A.~Salgado,
  Eur.\ Phys.\ J.\ C {\bf 77}, no. 3, 163 (2017)



\bibitem{Lokhtin:2005px}
  I.~P.~Lokhtin and A.~M.~Snigirev,
  Eur.\ Phys.\ J.\ C {\bf 45}, 211 (2006)


\bibitem{Lokhtin:2006qm}
  I.~P.~Lokhtin and A.~M.~Snigirev,
  J.\ Phys.\ G {\bf 34}, S999 (2007)




\bibitem{Lokhtin:2011qq}
  I.~P.~Lokhtin, A.~V.~Belyaev and A.~M.~Snigirev,
  Eur.\ Phys.\ J.\ C {\bf 71}, 1650 (2011)





\bibitem{Lokhtin:2014vda}
  I.~P.~Lokhtin, A.~A.~Alkin and A.~M.~Snigirev,
  Eur.\ Phys.\ J.\ C {\bf 75}, no. 9, 452 (2015)

\bibitem{Zapp:2008gi}
  K.~Zapp, G.~Ingelman, J.~Rathsman, J.~Stachel and U.~A.~Wiedemann,
  Eur.\ Phys.\ J.\ C {\bf 60}, 617 (2009)


\bibitem{Zapp:2011ya}
  K.~C.~Zapp, J.~Stachel and U.~A.~Wiedemann,
  JHEP {\bf 1107}, 118 (2011)

\bibitem{KunnawalkamElayavalli:2016ttl}
  R.~Kunnawalkam Elayavalli and K.~C.~Zapp,
  Eur.\ Phys.\ J.\ C {\bf 76}, no. 12, 695 (2016)
  doi:10.1140/epjc/s10052-016-4534-6
  [arXiv:1608.03099 [hep-ph]].

\bibitem{KunnawalkamElayavalli:2017hxo}
  R.~K. Elayavalli and K.~C.~Zapp,
  JHEP {\bf 1707}, 141 (2017)




\end{thebibliography}
\end{document}